\documentclass{article}
\pdfoutput=1

\PassOptionsToPackage{numbers, compress}{natbib}
\usepackage[preprint]{nips_2018}

\usepackage[utf8]{inputenc} % allow utf-8 input
\usepackage[T1]{fontenc}    % use 8-bit T1 fonts
\usepackage{hyperref}       % hyperlinks
\usepackage{url}            % simple URL typesetting
\usepackage{booktabs}       % professional-quality tables
\usepackage{amsfonts}       % blackboard math symbols
\usepackage{nicefrac}       % compact symbols for 1/2, etc.
\usepackage{microtype}      % microtypography

\usepackage[final]{graphicx}
\usepackage{epstopdf}
\graphicspath{ {Figures/} } 

\usepackage{multirow}

\usepackage{amsmath}

\usepackage{collcell,array}
\newcommand{\shiftdown}[1]{\smash{\raisebox{2.5\normalbaselineskip}{#1}}}
\newcolumntype{C}{>{\collectcell\shiftdown}c<{\endcollectcell}}

\title{Convolutional neural networks with \\ extra-classical receptive fields}

\author{
  Brian Hu \\
  Allen Institute for Brain Science\\
  Seattle, WA 98109 \\
  \texttt{brianh@alleninstitute.org} \\
  \And
  Stefan Mihalas \\
  Allen Institute for Brain Science\\
  Seattle, WA 98109 \\
  \texttt{stefanm@alleninstitute.org} \\
}

\begin{document}

\maketitle

\begin{abstract}
Convolutional neural networks (CNNs) have had great success in many real-world applications and have also been used to model visual processing in the brain. However, these networks are quite brittle -- small changes in the input image can dramatically change a network's output prediction. In contrast to what is known from biology, these networks largely rely on feedforward connections, ignoring the influence of recurrent connections. They also focus on supervised rather than unsupervised learning.
% take out rhetorical bit
%Can we use knowledge from biology to improve the robustness of these systems by including recurrent lateral connections learned in an unsupervised manner?
To address these issues, we combine traditional
supervised learning via backpropagation %to learn different features within a convolutional neural network, 
with a specialized
unsupervised learning rule to learn lateral connections between neurons within a convolutional neural network. %the neurons encoding these features.
These connections have been shown to optimally integrate information from the surround, generating extra-classical receptive fields for the neurons in our new proposed model (CNNEx).
%These connections are biologically plausible, and can be implemented using a collection of cell types present in cortical circuits.
Models with optimal lateral connections are more robust to noise and achieve better performance on noisy versions of the MNIST and CIFAR-10 datasets. Resistance to noise can be further improved by combining our model with additional regularization techniques such as dropout and weight decay. Although the image statistics of MNIST and CIFAR-10 differ greatly, the same unsupervised learning rule generalized to both datasets. Our results demonstrate the potential usefulness of combining supervised and unsupervised learning techniques %in real-world vision tasks
and suggest that the integration of lateral connections into convolutional neural networks is an important area of future research.
\end{abstract}

\section{Introduction}

% The notion of the extra-classical receptive field is not very strong here...
The visual response of a neuron is traditionally characterized by its classical receptive field (RF). However, such a picture of response tuning is incomplete as neurons can also integrate contextual information from other sources such as their surround. Contextual modulation refers to the ability of visual stimuli far outside the classical RF of a neuron to modulate the activity of the neuron. Examples of contextual modulation include surround suppression~\cite{Bair_etal03,Jones_etal01}, contour integration~\cite{Hess_etal03,Li_etal06}, and figure-ground segmentation~\cite{Zhou_etal00,Lamme_etal95}. These phenomena cannot be simply explained by feedforward mechanisms, but instead suggest an influence from extra-classical RFs. Contextual modulation is thought to be mediated in part by lateral connections between neurons in the same visual area~\cite{Angelucci_Bressloff06}. Experimental and computational modeling studies suggest that both excitatory and inhibitory cell types play an important role in this process~\cite{Ko_etal11,Jiang_etal15,Lee_etal16}.

% Need a better segue?
The field of deep learning has traditionally focused on feedforward models of visual processing, and these models have been used to describe neural activity in the ventral stream of humans and other primates~\cite{Cadieu_etal14, Gucclu_vanGerven15, Yamins_Dicarlo16, Wang_Cottrell17}.
% Add reference to work in mice as well, e.g. Brain Obs paper?
Deep learning is a form of supervised learning that relies on backpropagation of a global error signal, but whether the brain actually uses backpropagation is controversial given the requirement for large amounts of labeled data and non-local updates to synaptic weights (for recent reviews on the connection between deep learning and neuroscience, see~\cite{Marblestone_etal16,Kietzmann_etal17}). While convolutional neural networks have resulted in many practical successes~\cite{Gu_etal17}, they can be highly susceptible to adversarial examples. In one extreme case, the change of a single pixel within the input image can with high confidence change the output prediction of the network~\cite{Su_etal17}. In contrast, visual processing in the brain makes use of recurrent connections, including top-down and lateral connections, which may provide some level of immunity to these adversarial attacks (for recent results on human adversarial examples, see~\cite{Elsayed_etal18}). More recently, convolutional neural networks that include recurrent connections have also been proposed~\cite{Spoerer_etal17}.

\begin{figure}[t]
  \centering 
  \includegraphics[width=0.6\textwidth]{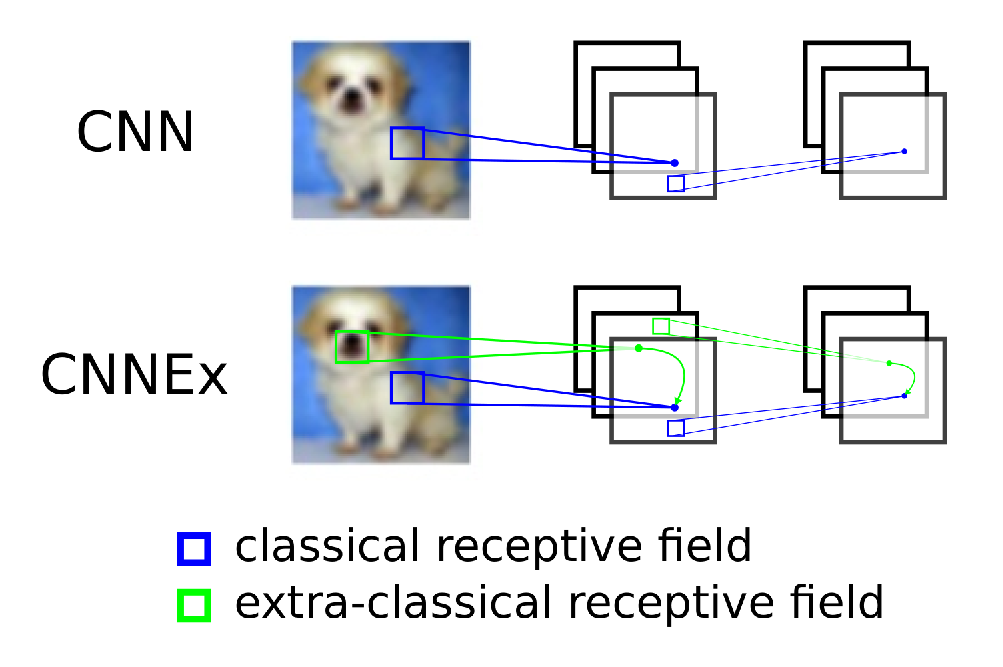}
  \caption{Schematic of network architectures with feedforward classical receptive fields and with added extra-classical receptive fields, termed CNN (top) and CNNEx (bottom), respectively. The classical receptive field of a unit in the network is shown in blue and an extra-classical receptive field is shown in green. In CNNEx, neurons from extra-classical receptive fields can modulate the activity of other neurons within a layer via recurrent lateral connections learned in an unsupervised manner. Here, only the two convolutional layers of the network are shown for simplicity. The input image is reproduced from the CIFAR-10 dataset~\protect{\cite{CIFAR}}.}
  \label{fig:FGM}
\end{figure}

However, most of these models still largely rely on supervised learning. The brain is able to build rich internal representations of information with little to no labeled data, which is a form of unsupervised learning. Recent work proposed Bayes optimal context integration as a canonical cortical computation, showing that optimal lateral connections can be learned using a modified Hebbian learning rule~\cite{Iyer_Mihalas17}. We extend this work by incorporating these types of lateral connections learned in an unsupervised manner into convolutional neural networks, which are trained in a supervised manner. We first train convolutional neural networks using standard backpropagation techniques. After training, we learn the optimal lateral connections between neurons within a layer in an unsupervised manner. We then test our models on two standard computer vision datasets, MNIST~\cite{MNIST} and CIFAR-10~\cite{CIFAR}. When applying different noise perturbations to the input images, the optimal lateral connections improve the overall performance and robustness of these networks. Our results suggest that incorporating lateral connections within convolutional neural networks is an important area of future research.
% Show images with perturbations
\begin{figure}[h]
\centering
\begin{tabular}{c c c}
\raisebox{28pt}[]{Original} & \hspace*{-12pt}\includegraphics[width=0.45\textwidth]{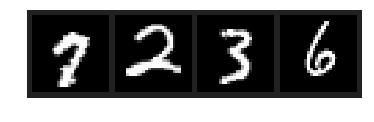}
& \hspace*{-12pt}\includegraphics[width=0.45\textwidth]{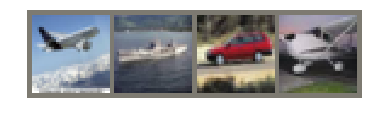}\\[-15pt]
\raisebox{28pt}[]{AWGN} & \hspace*{-12pt}\includegraphics[width=0.45\textwidth]{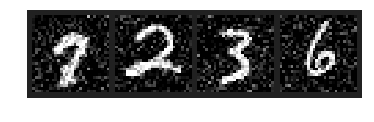}
& \hspace*{-12pt}\includegraphics[width=0.45\textwidth]{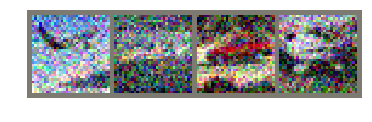}\\[-15pt]
\raisebox{28pt}[]{SPN} & \hspace*{-12pt}\includegraphics[width=0.45\textwidth]{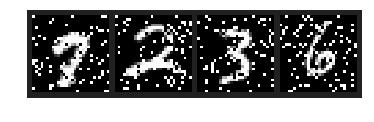}
& \hspace*{-12pt}\includegraphics[width=0.45\textwidth]{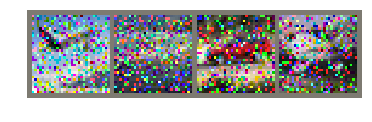}\\[-10pt]
& (A) MNIST & (B) CIFAR-10
\end{tabular}
\makeatletter
\let\@currsize\normalsize
\caption{Image datasets used in the experiments, MNIST (A) and CIFAR-10 (B). Along with the original images, we introduced two types of noise perturbations: additive white gaussian noise (AWGN) and salt-and-pepper noise (SPN). The top row shows the original images, the middle row shows the AWGN stimuli, and the bottom row shows the SPN stimuli. For the example noise stimuli shown here, the AWGN stimuli had a mean of zero and a standard deviation of 0.2 and the SPN stimuli had a fraction of changed pixels set to 0.2. The original images are reproduced from the MNIST~\protect{\cite{MNIST}} and CIFAR-10~\protect{\citep{CIFAR}} datasets.}
\label{fig:Datasets}
\end{figure}

\section{Methods}

\subsection{Image datasets: MNIST and CIFAR-10}
We trained and evaluated our models on two image datasets: MNIST~\cite{MNIST} and CIFAR-10~\cite{CIFAR}. MNIST contains grayscale images (28x28 pixels) of handwritten digits (10 classes, for the digits 0-9). MNIST contains a total of 70K images, split into a training set (60K images) and a test set (10K images). We used 10\% of the training data (6K images) for validation. CIFAR-10 contains color images (32x32 pixels) of objects from ten different classes (e.g. car, ship, etc.) CIFAR-10 contains a total of 60K images, again split into a training set (50K images) and a test set (10K images). We again used 10\% of the training data (5K images) for validation.

To test the generalization of our models under noise perturbations, we added two types of noise to the original images: additive white Gaussian noise (AWGN) and salt-and-pepper noise (SPN). The mean of the AWGN was set to zero and the standard deviation varied in increasing levels of $\{0.1, 0.2, 0.3, 0.4, 0.5\}$. For the SPN, the fraction of noisy pixels varied in increasing levels of $\{0.1, 0.2, 0.3, 0.4, 0.5\}$.
% Added- not sure if necessary or not
The addition of noise can be viewed as a random, non-targeted adversarial attack, which changes the input image in such a way that it will be classified incorrectly. The degree of misclassification is dependent on the noise level.
Example stimuli from each dataset (original and noisy images) are shown in Figure~\ref{fig:Datasets}.

\subsection{Network architecture and training}
We used a simple network architecture to study the influence of optimal lateral connections in convolutional neural networks. The baseline network consisted of two convolutional (conv) layers with the ReLU nonlinearity, each followed by a max-pooling (maxpool) layer with a 2x2 pooling window, which effectively downsamples the input by a factor of 2. Following the two convolutional layers are two fully connected (FC) layers, with the final output passed through a soft-max nonlinearity for the 10 classes in each dataset. We used the same baseline model architecture for both MNIST and CIFAR-10 (Table~\ref{network_structure}). We also used the same set of hyperparameters for training both models, namely stochastic gradient descent with a learning rate of
% Try increasing momentum to 0.9? 0.5 is somewhat nonstandard
% Try varying architecture/training epochs for CIFAR-10
0.01 and a momentum value of 0.5. We used a minibatch size of 64 and trained our models for a total of 10 epochs. We trained 10 different instantiations of each model using different random seeds to ensure the robustness of our results. All experiments were performed using Pytorch (0.3.1) on a NVIDIA GTX 1080 Ti GPU.

% Show table of model architecture
\begin{table}[b]
  \caption{Model architectures used for the experiments. CNN is the baseline model without lateral connections, and CNNEx is the model with optimal lateral connections. Convolutional layers are denoted as ``conv$<$receptive field size$>$-$<$number of channels$>$''. Convolutional layers in italics represent recurrent lateral connections learned in an unsupervised manner. ``maxpool'' denotes max pooling using a 2x2 window and a stride of 2. ``FC'' denotes fully connected layers with the given number of units. The ReLU activation function is not shown for brevity.}
%  The number of parameters of the two models were matched as closely as possible.}
  \label{network_structure}
  \centering
  \resizebox{\textwidth}{!}{
  \begin{tabular}{cccccccccc}
    \toprule
    % report number of params for each model as well?
    \textbf{Model} & \multicolumn{9}{c}{\textbf{Network architecture}}                   \\
    \midrule
    % conv5-13 and conv5-26
    CNN & \multicolumn{2}{c}{conv5-10} & maxpool & \multicolumn{2}{c}{conv5-20} & maxpool & FC-50 & FC-10 & soft-max \\
    \midrule
    CNNEx & conv5-10 & \textit{conv7-10} & maxpool & conv5-20 & \textit{conv3-20} & maxpool & FC-50 & FC-10 & soft-max \\
%         \midrule
%      \multirow{2}{*}{CIFAR-10} & CNN & 42K & \multicolumn{2}{c}{conv5-13} & maxpool & \multicolumn{2}{c}{conv5-26} & maxpool & FC-50 & FC-10 & soft-max \\
%     \cmidrule(lr){2-12}
%     & CNNEx & 40K & conv5-10 & \textit{conv7-10} & maxpool & conv5-20 & \textit{conv3-20} & maxpool & FC-50 & FC-10 & soft-max \\
    \bottomrule
  \end{tabular}
  }
\end{table}

\subsection{Optimal lateral connections}
After the initial phase of supervised learning, we freeze the feedforward synaptic weights of the network. The classical receptive field response of a neuron representing feature $j$ in layer $l$ at image location $m$, given image $x$ can be represented by the activation of a standard artificial neuron model:
% Give equations here, can expand the math if needed
\begin{equation}
fc^{m,l}_{j,x} = \phi \left( b + \sum_k\sum_{n} U^{mn}_{jk}fc^{n,l-1}_{k,x} \right)
\label{eq1}
\end{equation}
where $\phi$ represents a nonlinear activation function, $b$ represents a bias term, $k$ represents features, $n$ represents image locations, and $U^{mn}_{jk}$ are the feedforward synaptic weights from layer $l-1$ to layer $l$.

We then apply optimal lateral connections within the first two convolutional layers of the network.
% drop the l superscript in the feedforward equation
Here, we drop the $l$ superscript, since the proposed lateral connections are intracortical and occur within the same layer. The lateral connections are between neurons with the same feature (within-channel) and neurons with different features (between-channel) over a fixed spatial extent.
The activity of a neuron representing feature $j$ at image location $m$, given image $x$ can then be written as:

% Give equations here, can expand the math if needed
\begin{equation}
% f^m_{j,x} = \frac{1}{p^m_x} fc^{m}_{j,x} \Big (1+\alpha\sum_k\sum^N_{n\neq m} W^{mn}_{jk}fc^{n}_{k,x} \Big )
f^m_{j,x} = fc^{m}_{j,x} \Big (1+\alpha\sum_k\sum_{n\neq m} W^{mn}_{jk}fc^{n}_{k,x} \Big )
\label{eq1}
\end{equation}
% where $p^m_x$ represents a normalization coefficient for patch $m$ in image $x$, 
where $f^{m}_{j,x}$ represents the full response of the neuron with contributions from extra-classical receptive fields, $fc^{m}_{j,x}$ represents the classical receptive field response of the neuron, $\alpha$ represents a hyperparameter that tunes the strength of the lateral connections, and $W^{mn}_{jk}$ are the synaptic weights from surrounding neurons. The lateral connections have a modulatory effect on the feedforward response, and %. As a result, 
setting $\alpha=0$ is equivalent to a model with no lateral connections.
% % This statement is a bit confusing, so I'm taking it out
% A hyperparameter for the lateral connections is needed as multiple nearby patches can provide redundant information about the same pixels in the image.

The synaptic weights are learned in an unsupervised manner using the following rule:

\begin{equation}
W^{mn}_{jk} = \frac{\Big \langle fc^{m}_{j} fc^{n}_{k} \Big \rangle_x}{\Big \langle fc^{m}_{j} \Big \rangle_x \Big \langle fc^{n}_{k} \Big \rangle_x} - 1
\label{eq2}
\end{equation}
where $W^{mn}_{jk}$ is the synaptic weight between each pair $(j,k)$ of features located at $(m,n)$ and $x$ spans a set of images. We used the same set of training images originally shown to the network during supervised training to learn the optimal lateral connections. However, for this phase of learning, we do not need the image labels, as our method is completely unsupervised. It is important to note that this formula differs from a Hebbian learning rule, in that only the covariance between the feedforward responses of neurons leads to changes in the lateral connections. A more detailed derivation of the above equations can be found in~\cite{Iyer_Mihalas17}.

\subsection{Network regularization}
To understand how other commonly used regularization techniques could impact model performance on noisy images, we also tested the use of weight decay ($L_2$ regularization) and dropout. For our experiments, we chose a weight decay value of 0.005 and a dropout fraction of 0.5. Weight decay acted on all non-bias parameters of the model, while dropout was applied after each convolutional layer in the baseline model, as well as after the first fully connected layer. We also tested the combination of these regularization techniques with optimal lateral connections.

\subsection{Validation and testing}
% change to 5x5 if this is the case
Optimal lateral connections had a spatial extent of 7x7 pixels in the first convolutional layer and 3x3 pixels in the second convolutional layer. We did not include any self-connections, so these were all set to zero. We chose the optimal lateral connection hyperparameters $\alpha$ for each of the two convolutional layers based on a coarse grid search over the parameter range $\{0.1, 0.01, 0.001, 0.0001\}$ using the validation dataset. 
% For our final results, we chose $\alpha=0.01$ for the first convolutional layer and $\alpha=0.1$ for the second convolutional layer.
We did not use lateral connections for the two fully-connected layers. We report final accuracies of each model on the original dataset and for all levels of the two different types of noise perturbations. All final results are averages over each of the 10 pre-trained models with different random seeds.
\begin{figure}[t]
\centering
\begin{tabular}{c}
\includegraphics[width=0.8\textwidth]{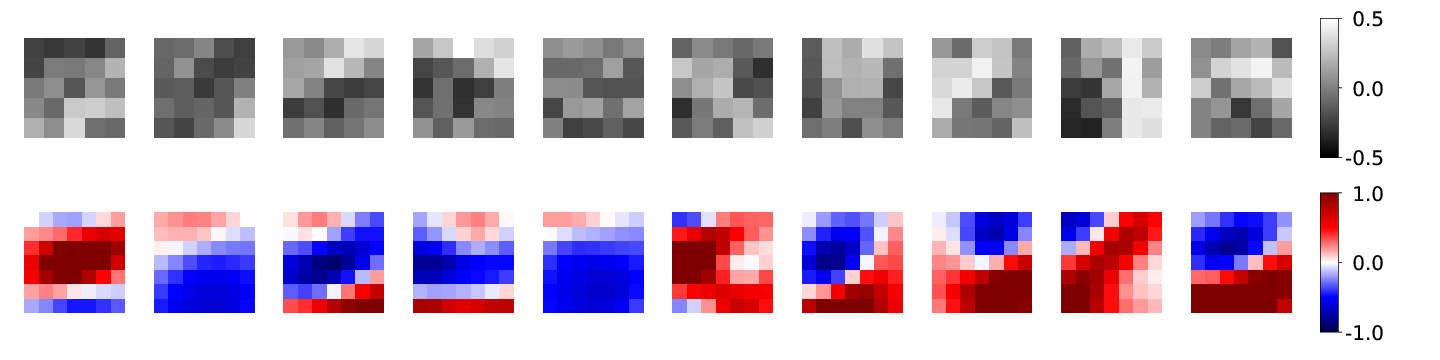}\\
(A) MNIST \\
\includegraphics[width=0.8\textwidth]{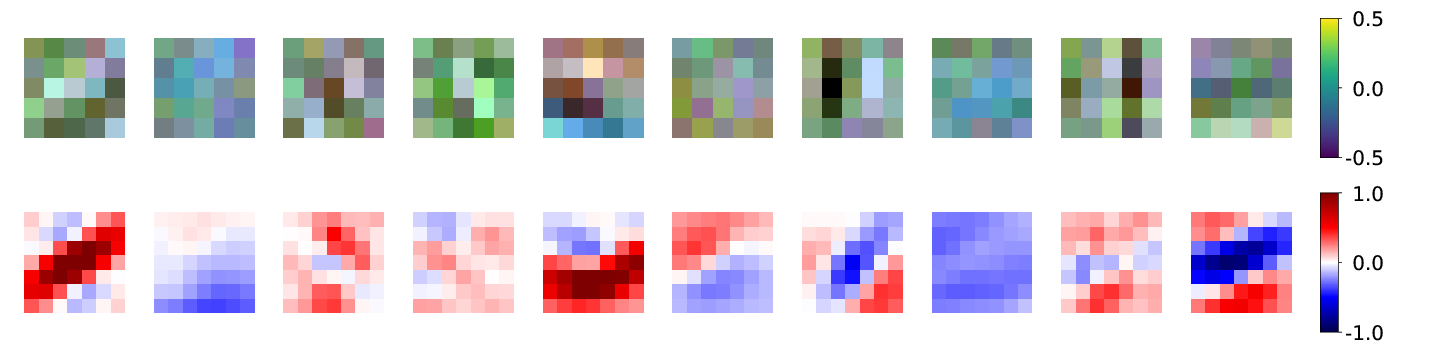}\\
(B) CIFAR-10
\end{tabular}
\makeatletter
\let\@currsize\normalsize
\caption{Example optimal lateral connections learned on the MNIST (A) and CIFAR-10 (B) datasets. The first row shows filters for the first convolutional layer learned in a supervised manner. The second row shows optimal lateral connections from each filter onto the first filter in each row learned in an unsupervised manner. The learned filters were 5x5 and the optimal lateral connections were 7x7.} % change to 5x5 if this is the case
\label{fig:OptimalConnections}
\end{figure}

\section{Results}

\subsection{Learned optimal lateral connections}

We show example learned optimal lateral connections between different filters in the first convolutional layer for both MNIST and CIFAR-10 (Figure~\ref{fig:OptimalConnections}). For both datasets, we find optimal lateral connections containing both excitatory and inhibitory synaptic weights, which are balanced on average. In particular, we find excitatory weights between cells with similar tuning properties and inhibitory weights between cells with different selectivities, which is consistent with predictions from Hebbian plasticity. Interestingly, although the image statistics between MNIST and CIFAR-10 are vastly different, some of the learned optimal connections are qualitatively similar, emphasizing properties such as contour integration which may be beneficial in the context of noise or occlusion.

\subsection{Accuracy on deep learning models}
% Table of results here:
\begin{table}
  \caption{Model accuracy (\%) on the MNIST and CIFAR-10 datasets. We separate results for the original images and the two types of noise perturbations by columns (AWGN: additive white gaussian noise, SPN: salt-and-pepper noise). The results for the baseline model (CNN) and the model with optimal lateral connections (CNNEx) are shown in separate rows. $wd$ corresponds to models trained with weight decay = 0.005. $d$ corresponds to models trained with a dropout fraction of 0.5. All reported values are averages over 10 random initializations.}
  \label{accuracy_table}
  \centering
  \resizebox{\textwidth}{!}{
\begin{tabular}{ccccccccccccc}
    \toprule
    & \multicolumn{1}{c}{\multirow{2}{*}{\textbf{Models}}} & \multicolumn{1}{c}{\textbf{Original}} & \multicolumn{5}{c}{\textbf{AWGN}} & \multicolumn{5}{c}{\textbf{SPN}} \\
    \cmidrule(lr){3-3}\cmidrule(lr){4-8}\cmidrule(lr){9-13}
    & & - & 0.1 & 0.2 & 0.3 & 0.4 & 0.5 & 0.1 & 0.2 & 0.3 & 0.4 & 0.5 \\
    \midrule
    % change all entries to CNN/CNNEx since the nomenclature was defined in Table 1
    %%% MNIST %%%
     \multirow{9}{*}{MNIST} & CNN & \textbf{98.77} & \textbf{98.68} & \textbf{98.31} & 96.87 & 91.58 & 80.96 & 97.40 & 92.00 & 80.04 & 64.00 & 46.94 \\
    % % original CNN
    %  \multirow{9}{*}{MNIST} & CNN & 98.77 & \textbf{98.69} & 98.30 & \textbf{96.80} & 91.00 & 80.54 & \textbf{97.42} & 91.66 & 79.79 & 63.91 & 46.85\\
    & CNNEx & 97.13 & 97.10 & 96.81 & 95.96 & 93.73 & 88.95 & 96.09 & 93.92 & 88.61 & 78.55 & 63.30 \\
    % % original CNNEx
    %  & CNNEx & 97.99 & 97.90 & 97.54 & 96.45 & 93.50 & 86.79 & 97.05 & 94.08 & 86.63 & 73.19 & 56.18 \\

     % original CNN with dropout
%      & CNN-Dropout & 98.26 & 98.21 & 97.97 & 97.33 & 95.85 & 92.64 & 97.54 & 95.73 & 91.21 & 82.66 & 70.01\\
     % original CNNEx with dropout
%      & CNNEx-Dropout & 98.13 & 98.07 & 97.84 & 97.28 & 95.96 & 93.17 & 97.40 & 95.71 & 91.91 & 84.69 & 72.04\\

    % % baseline matched params
    %  \multirow{9}{*}{MNIST} & CNN & \textbf{98.77} & \textbf{98.67} & \textbf{98.32} & 96.72 & 91.33 & 80.93 & 97.21 & 91.53 & 79.85 & 64.00 & 47.00\\
    % % CNNEx (with fine-tuning)
    % & CNNEx & 98.59 & 98.53 & 98.16 & 97.01 & 92.75 & 83.42 & 97.55 & 93.54 & 83.45 & 68.40 & 50.76 \\

	% supervised lateral (performs the best)
%     & CNN-S & 98.74 & 98.66 & 98.32 & 97.35 & 94.48 & 87.91 & 97.69 & 94.57 & 87.43 & 75.12 & 59.46\\

	\cmidrule(lr){2-13}
	% CNN with weight decay
     & CNN (wd) & 98.57 & 98.48 & 98.18 & \textbf{97.37} & 94.48 & 87.21 & \textbf{97.70} & 95.08 & 87.25 & 73.57 & 56.25 \\
    %  & CNN (wd) & 98.49 & 98.41 & 98.11 & 97.24 & 94.37 & 87.26 & 97.64 & 94.81 & 87.00 & 73.31 & 56.68\\
    % baseline matched params (weight decay)
    %  & CNN (wd) & 98.65 & 98.55 & 98.29 & 97.46 & 94.41 & 87.13 & \textbf{97.82} & 94.95 & 87.05 & 73.82 & 57.01\\
    % CNNEx with weight decay
     & CNNEx (wd) & 97.53 & 97.44 & 97.17 & 96.50 & 94.93 & 91.55 & 96.75 & \textbf{95.10} & 91.49 & 84.01 & 70.58 \\
    % % CNNEx (with fine-tuning)
    %   & CNNEx (wd) & 98.16 & 98.11 & 97.85 & \textbf{97.16} & 94.93 & 89.98 & 97.46 & 95.46 & 90.25 & 80.26 & 65.38 \\
     
     % dropout
     	\cmidrule(lr){2-13}
      & CNN (d) & 97.44 & 97.34 & 96.96 & 96.25 & 94.47 & 90.48 & 96.24 & 94.00 & 88.86 & 79.86 & 66.08 \\
    % % baseline matched params (dropout)
    %  & \textit{CNN} (d) & 97.96 & 97.86 & 97.52 & 96.85 & 95.27 & 91.78 & 96.95 & 94.95 & 90.56 & 82.29 & 68.88 \\
     % CNNEx with dropout (no dropout in FC layers)
    %  & CNNEx (d) & 98.13 & 98.07 & 97.84 & 97.28 & 95.96 & 93.17 & 97.40 & 95.71 & 91.91 & 84.69 & 72.04 \\
     & CNNEx (d) & 96.63 & 96.61 & 96.36 & 95.80 & 94.74 & 92.65 & 95.92 & 94.45 &
       91.86 & 86.57 & 76.64 \\
    %  % with fine-tuning
    %   & CNNEx (d) & 97.21 & 97.13 & 96.85 & 96.19 & 94.80 & 92.10 & 96.36 & 94.70 & 91.24 & 84.29 & 71.94 \\
     
    % weight decay + dropout
    \cmidrule(lr){2-13}
     & CNN (wd+d) & 97.22 & 97.07 & 96.79 & 96.24 & 95.02 & 92.43 & 96.46 & 94.89 & 91.55 & 84.71 & 72.71 \\
    % % baseline matched params (weight decay)
    %  & \textit{CNN (wd+d)} & 97.68 & 97.59 & 97.33 & 96.82 & \textbf{95.75} & \textbf{93.40} & 97.01 & \textbf{95.69} & 92.84 & 86.65 & 75.21 \\
     & CNNEx (wd+d) & 96.97 & 96.88 & 96.65 & 96.12 & \textbf{95.14} & \textbf{93.37} & 96.32 & 95.02 & \textbf{92.88} & \textbf{88.25} & \textbf{78.93} \\
    % % with fine-tuning
    %   & CNNEx (wd+d) & 97.16 & 97.11 & 96.78 & 96.26 & 95.25 & 93.24 & 96.51 & 95.27 & \textbf{92.92} & \textbf{87.64} & \textbf{77.44} \\

     \midrule
     %% CIFAR-10
     \multirow{9}{*}{CIFAR-10} & CNN & \textbf{59.60} & 51.53 & 36.56 & 26.00 & 20.59 & 17.68 & 37.87 & 26.11 &
       20.31 & 17.23 & 15.28 \\
    % % original CNN
    % \multirow{9}{*}{CIFAR-10} & CNN & 60.10 & 51.98 & 37.94 & 28.71 & 23.36 & 20.07 & 39.28 & 29.11 & 23.33 & 19.55 & 16.89 \\
     & CNNEx & 58.04 & \textbf{52.20} & 38.97 & 28.34 & 22.25 & 18.86 & 40.17 & 28.38 & 21.96 & 18.18 & 15.94 \\
% % 	original CNNEx
%      & CNNEx & 59.23 & \textbf{52.93} & 39.50 & 29.61 & 23.67 & 20.21 & 41.16 & 30.11 & 23.75 & 19.70 & 16.97 \\

    % CNN with dropout
%      & CNN-Dropout & 47.49 & 45.92 & 41.90 & 37.03 & 31.92 & 27.67 & 42.30 & 36.70 & 31.38 & 26.46 & 22.38\\
%      CNNEx with dropout
%      & CNNEx-Dropout & 47.67 & 46.07 & 42.09 & 37.00 & 31.96 & 27.58 & 42.62 & 36.80 & 31.53 & 26.48 & 22.22\\

    % % baseline matched params
    % \multirow{9}{*}{CIFAR-10}  & CNN & 62.14 & 55.01 & 40.00 & 28.92 & 22.57 & 19.05 & 40.91 & 28.66 & 22.11 & 18.23 & 15.76 \\
    %  % 	CNNEx (with fine-tuning)
    %  & CNNEx & \textbf{63.73} & 57.62 & 41.93 & 29.86 & 22.88 & 19.11 & 43.21 & 29.85 & 22.70 & 18.57 & 16.03 \\
     
% 	% supervised lateral (performs the best)
%     & CNN-S & 57.20 & 53.00 & 42.58 & 32.21 & 24.81 & 20.24 & 43.97 & 32.90 & 24.99 & 19.36 & 15.69 \\

	\cmidrule(lr){2-13}
     & CNN (wd) & 57.62 & 51.41 & 38.68 & 28.15 & 21.94 & 18.39 & 40.02 & 28.26 & 21.63 & 17.79 & 15.57 \\
    % CNN with weight decay
    %  & CNN (wd) & 57.90 & 52.14 & 39.64 & 29.37 & 23.43 & 20.10 & 41.06 & 29.40 & 23.09 & 19.29 & 16.98 \\
    %   % baseline matched params (weight decay)
    %  & CNN (wd) & 59.95 & 54.46 & 41.73 & 31.09 & 24.40 & 20.38 & 42.77 & 30.71 & 23.79 & 19.39 & 16.70 \\
%     CNNEx with weight decay
     & CNNEx (wd) & 56.57 & 51.89 & \textbf{40.85} & 30.61 & 23.95 & 19.95 & \textbf{42.20} & 30.83 & 23.67 & 19.16 & 16.48 \\
    % % with fine-tuning
    %   & CNNEx (wd) & 62.92 & \textbf{57.94} & \textbf{44.93} & 32.87 & 25.32 & 20.97 & \textbf{46.15} & 32.93 & 25.00 & 20.05 & 17.19 \\
     
     % dropout
     	\cmidrule(lr){2-13}
      & CNN (d) & 43.22 & 42.12 & 38.95 & 34.22 & 29.50 & 25.64 & 39.30 & 34.30 & 29.29 & 24.70 & 20.84 \\
          % % baseline matched params (dropout)
    %  & \textit{CNN} (d) & 45.96 & 44.80 & 40.96 & 35.53 & 30.13 & 25.57 & 41.38 & 35.66 & 29.81 & 24.43 & 20.35 \\
%      CNNEx with dropout (no FC layers)
    %  & CNNEx (d) & 47.67 & 46.07 & 42.09 & 37.00 & 31.96 & 27.58 & 42.62 & 36.80 & 31.53 & 26.48 & 22.22\\
      & CNNEx (d) & 43.68 & 42.51 & 39.20 & 34.36 & 29.65 & 25.76 & 39.70 & 34.57 & 29.55 & 24.78 & 20.85 \\
    % % with fine-tuning
    % & CNNEx (d) & 43.85 & 42.53 & 38.67 & 33.27 & 28.05 & 23.85 & 39.16 & 33.41 & 28.06 & 22.87 & 18.79 \\
     
     % weight decay + dropout
    \cmidrule(lr){2-13}
     & CNN (wd+d) & 43.05 & 42.52 & 40.35 & \textbf{36.87} & \textbf{32.55} & \textbf{28.85} & 40.64 & 37.14 & \textbf{32.87} & \textbf{28.07} & \textbf{23.47} \\
    % % baseline matched params (weight decay)
    %  & \textit{CNN} (wd+d) & 45.70 & 44.88 & 42.68 & \textbf{39.17} & \textbf{34.95} & \textbf{30.71} & 42.92 & \textbf{39.20} & \textbf{34.80} & \textbf{29.67} & \textbf{24.62} \\
    %   CNNEx with dropout
     & CNNEx (wd+d) & 43.12 & 42.54 & 40.38 & \textbf{36.87} & 32.53 & 28.83 & 40.67 & \textbf{37.15} & 32.86 & 28.05 & 23.46 \\
    % % with fine-tuning
    %   & CNNEx (wd+d) & 43.46 & 42.58 & 40.03 & 36.01 & 31.29 & 27.15 & 40.52 & 36.42 & 31.67 & 26.23 & 21.61 \\
     
    \bottomrule
  \end{tabular}
  }
\end{table}

%     & CNN & \textbf{98.77} & \textbf{98.69} & \textbf{98.30} & \textbf{96.80} & 91.00 & 80.54 & \textbf{97.42} & 91.66 & 79.79 & 63.91 & 46.85\\
%     \multirow{1}{*}{MNIST} & CNNEx & 97.99 & 97.90 & 97.54 & 96.45 & \textbf{93.50} & \textbf{86.79} & 97.05 & \textbf{94.08} & \textbf{86.63} & \textbf{73.19} & \textbf{56.18} \\
% %     \cmidrule(lr){2-13}
% %     & CNN-Dropout & \textbf{98.26} & \textbf{98.21} & \textbf{97.97} & \textbf{97.33} & 95.85 & 92.64 & \textbf{97.54} & \textbf{95.73} & 91.21 & 82.66 & 70.01\\
% % %     & CNN-Control-Dropout & \\
% %     & CNNEx-Dropout & 98.13 & 98.07 & 97.84 & 97.28 & \textbf{95.96} & \textbf{93.17} & 97.40 & 95.71 & \textbf{91.91} & \textbf{84.69} & \textbf{72.04}\\
%     \midrule
%     & CNN & \textbf{60.10} & 51.98 & 37.94 & 28.71 & 23.36 & 20.07 & 39.28 & 29.11 & 23.33 & 19.55 & 16.89 \\
%     \multirow{1}{*}{CIFAR-10} & CNNEx & 59.23 & \textbf{52.93} & \textbf{39.50} & \textbf{29.61} & \textbf{23.67} & \textbf{20.21} & \textbf{41.16} & \textbf{30.11} & \textbf{23.75} & \textbf{19.70} & \textbf{16.97} \\

% Explain performance improvement here (refactor based on new results)
We tested our trained models with and without optimal lateral connections on the original MNIST and CIFAR-10 test datasets, as well as on these datasets with the addition of noise. For the MNIST dataset, we find that both the baseline network and the network with optimal lateral connections achieve high accuracy on the original test images ($\sim$97-99\%). We also find that performance decreases gradually with increasing noise levels. In general, accuracy is lower for the SPN images compared to the AWGN images, suggesting that SPN images may be more difficult for the baseline model to handle. Our results show that optimal lateral connections improve model performance at higher levels of AWGN (standard deviations above 0.3) %4)
and also at higher levels of SPN (fraction of changed pixels above 0.1).
% In general, the optimal lateral connections seemed to improve performance of the models across the different noise types.
% added
Our results also show that the combination of optimal lateral connections with additional regularization techniques such as dropout or weight decay often resulted in even better performance at high noise levels.
% give more specific examples of performance boost
For example, at the highest noise levels, models with optimal lateral connections and regularized by both weight decay and dropout achieved the highest accuracy (93.37\% on AWGN images and 78.93\% on SPN images). Interestingly, the relative difference in accuracies between models with and without optimal lateral connections decreased with additional regularization (e.g. on the SPN images, $\sim$16\% difference for models without regularization and $\sim$6\% difference for models with weight decay and dropout).

For the CIFAR-10 dataset, the baseline model achieves an accuracy around 60\%. We find a slight decrease in the performance of the model with optimal lateral connections compared to the baseline model on the original set of images (which we also saw on the MNIST dataset). Models with optimal lateral connections again outperform the baseline models for both types of noise and at different noise levels. However, the increase in performance with optimal lateral connections is relatively small ($\sim$1-2\%) on the CIFAR-10 dataset compared to the MNIST dataset. We also find that for the model which is regularized by both weight decay and dropout, optimal lateral connections do not provide much additional benefit on the CIFAR-10 dataset. This may be due to our models being underfit, as even our best performing model does not achieve close to state-of-the-art test accuracies on the CIFAR-10 dataset. Our results are summarized in Table~\ref{accuracy_table}.
% Our results are summarized in Figure~\ref{fig:AccuracyResults}.

% Comment on combination of optimal lateral connections with other regularization techniques?

\section{Discussion}
% Please take some time to edit

% TODO: Comment on biological significance? Cell types that implement this computation?

% Think about supervised learning of lateral connections
% Think about unsupervised+supervise (e.g. ladder networks)
% Think about multiplicative weights...
% Think about connection to ResNets?

Our proposed model adds two novel contributions to traditional convolutional neural networks when considered together: 1) the incorporation of recurrent lateral connections modeling the influence of extra-classical receptive fields, and 2) the ability to learn these connections in a completely unsupervised manner.

The vast majority of deep neural networks are feedforward in nature, although recurrent connections have been added to convolutional neuronal networks~\cite{Spoerer_etal17,Liang_Hu15}. Recurrent connections have also been used to implement different visual attention mechanisms~\cite{Mnih_etal14,Li_etal17}. However, these networks are still all trained in a supervised manner. An exception are ladder networks, which have been proposed as a means to combine supervised and unsupervised learning in deep neural networks~\cite{Rasmus_etal15}. However, different from our approach, ladder networks use noise injection to introduce an unsupervised cost function based on reconstruction of the internal activity of the network. Our model instead relies on a modified Hebbian learning rule which learns the optimal connections between features within each layer based solely on the activations of these neurons.

% Comment on cell-type specific contributions in the circuit? Tie to Ram's paper?
%Strengths/predictions
The proposed computation carried out by optimal lateral connections can be mapped to a collection of cell types found in the cortical microcircuit~\cite{Jiang_etal15}. Connections between pyramidal cells often show like-to-like connectivity, and in our model, the strength of these connections is proportional to the correlation in the activations of these neurons. This mapping also suggests two forms of inhibition - local normalization of excitatory neuronal activity in a patch (corresponding to the classical receptive field) and inhibition arising from the surround (extra-classical receptive fields) - we attribute these to parvalbumin and somatostatin-expressing interneurons, respectively.

Neurons are inherently noisy, and their responses can vary even to the same stimulus. These neurons are embedded in cortical circuits that must perform computations in the absence of information, such as under visual occlusion. Optimal lateral connections can provide additional robustness to these networks by allowing for integration of information from multiple sources. This type of computation is also potentially useful for applications in which artificial neurons are not simulated with high fidelity, e.g. in neuromorphic computing.

% The use of additional regularization techniques such as dropout~\cite{srivastava2014dropout} may improve model performance under noise.
% % This may not be preliminary if it is also a row in the table
% Our preliminary tests show that even with models trained with dropout, the use of additional information from extra-classical receptive fields via optimal lateral connections can provide a performance boost
% 
% A future direction of research is to understand how to balance the strength of optimal lateral connections with biological constraints such as sparseness, which may provide additional forms of regularization. 

We chose a relatively simple network architecture as a proof-of-concept for our model. As such, we did not achieve state-of-the art performance on either image dataset. This accuracy could be further improved by either %incorporating data augmentation during training or 
fine-tuning models with optimal lateral connections or
using deeper model architectures with more parameters.
% Again, this may be done currently, so we may have to adjust accordingly
Future experiments will also have to test the scalability of learning optimal lateral connections on more complex network architectures and larger image datasets (e.g. ImageNet), and whether these connections provide any benefit against noise or other types of perturbations such as adversarial images.
%Not sure if we should give this away or not
%In our current model, we made a simplifying assumption that the learned lateral connections from extra-classical receptive fields have an additive effect in modulating the activity of a neuron. We are currently investigating the performance of models that do not make this assumption, and instead have extra-classical receptive fields which have a multiplicative contribution. This could be implemented in the biology by complex dendritic trees.
\subsection*{Acknowledgements}
We wish to thank the Allen Institute founder, Paul G. Allen, for his vision, encouragement, and support. We also thank Ram Iyer for helpful discussions.

\bibliographystyle{unsrt}
\bibliography{references.bib}

\end{document}